\documentclass[10pt,conference]{IEEEtran}

\usepackage{amsmath,amssymb,amsfonts}
\usepackage{cite}
\usepackage{graphicx}
\usepackage{epstopdf}

\usepackage{booktabs}
\usepackage{graphicx}
\usepackage{textcomp}
\usepackage{xcolor}
\usepackage{times}
\usepackage{subfigure}         
\usepackage{amssymb,amsmath}
\usepackage{acronym}  
\usepackage{balance}

\usepackage{bm}         
\usepackage{algorithm} 
\usepackage{algorithmic}

\usepackage{lipsum}
\usepackage{stfloats}

\usepackage{setspace}

\usepackage{multirow}

\usepackage{url}

\usepackage{mathrsfs}

\usepackage{amsmath}
\allowdisplaybreaks[4]

\newtheorem{definition}{\bf Definition}

\acrodef{siso}[SISO]{single-input single-output}%

\acrodef{nmse}[NMSE]{normalized mean square error}%

\acrodef{ris}[RIS]{reconfigurable intelligent surface}%

\acrodef{csi}[CSI]{channel state information}%
\acrodef{awgn}[AWGN]{additive white Gaussian noise}%

\acrodef{cs}[CS]{compressed sensing}%
\acrodef{ao}[AO]{alternaing optimization}%

\acrodef{bs}[BS]{base station}%
\acrodef{snr}[SNR]{signal-to-noise ratio}%
\acrodef{mmwave}[mmWave]{millimeter-wave}%
\acrodef{snr}[SNR]{signal-to-noise ratio}%
\acrodef{rf}[RF]{radio frequency}%

\acrodef{omp}[OMP]{orthogonal matching pursuit}%
\acrodef{ml}[ML]{maximum likelihood}%

\acrodef{mse}[MSE]{mean square error }%

\acrodef{sbr}[S-BAR]{successive Bayesian reconstructor}%

\acrodef{as}[AS]{antenna selection}%

\acrodef{fas}[FAS]{fluid antenna system}%

\acrodef{swipt}[SWIPT]{simultaneous wireless information and power transfer}%
\acrodef{sinr}[SINR]{signal-to-interference-plus-noise ratio}%

\acrodef{rc}[RC]{reflection coefficient}%
\acrodef{mimo}[MIMO]{multiple-input multiple-output}%

\def\BibTeX{{\rm B\kern-.05em{\sc i\kern-.025em b}\kern-.08em
		T\kern-.1667em\lower.7ex\hbox{E}\kern-.125emX}}

\begin{document}
	\title{Successive Bayesian Reconstructor for \\ FAS Channel Estimation \vspace*{-0.4em}}
	\author{
		\IEEEauthorblockN{
			Zijian~Zhang\IEEEauthorrefmark{1},~Jieao~Zhu\IEEEauthorrefmark{1},~Linglong~Dai\IEEEauthorrefmark{1},~{\it Fellow,~IEEE}, and Robert W. Heath Jr.\IEEEauthorrefmark{2}, {\it Fellow, IEEE}
		}
		\vspace{0.2cm}
		\IEEEauthorblockA{ \small
			\IEEEauthorrefmark{1}Department of Electronic Engineering, Tsinghua University, Beijing 100084, China \\
			\IEEEauthorrefmark{1}Beijing National Research Center for Information Science and Technology (BNRist), Beijing 100084, China \\	
			\IEEEauthorrefmark{2}Department of Electrical and Computer Engineering, University of California, San Diego, 9500 Gilman Drive, La Jolla, CA 92093, USA\\
			E-mails: \{zhangzj20, zja21\}@mails.tsinghua.edu.cn; daill@tsinghua.edu.cn; rwheathjr@ucsd.edu
			\vspace*{-2em}
		}
	}
	\maketitle
	
\begin{abstract}
Fluid antenna systems (FASs) can reconfigure their locations freely within a spatially continuous space. To keep favorable antenna positions, the channel state information (CSI) acquisition for FASs is essential. While some techniques have been proposed, most existing FAS channel estimators require several channel assumptions, such as slow variation and angular-domain sparsity. When these assumptions are not reasonable, the model mismatch may lead to unpredictable performance loss. In this paper, we propose the successive Bayesian reconstructor (S-BAR) as a general solution to estimate FAS channels. Unlike model-based estimators, the proposed S-BAR is prior-aided, which builds the experiential kernel for CSI acquisition. Inspired by Bayesian regression, the key idea of S-BAR is to model the FAS channels as a stochastic process, whose uncertainty can be successively eliminated by kernel-based sampling and regression. In this way, the predictive mean of the regressed stochastic process can be viewed as the maximum a posterior (MAP) estimator of FAS channels. Simulation results verify that, in both model-mismatched and model-matched cases, the proposed S-BAR can achieve higher estimation accuracy than the existing schemes.
\end{abstract}
	
\section{Introduction}	
In recent years, \acp{fas}, also called fluid antennas or movable antennas, are proposed to achieve higher diversity and multiplexing gains than conventional \ac{mimo} systems \cite{wong2020fluid,wong2021fluid,Wenyan'23}. Different from \ac{mimo} with fixed antennas, \ac{fas} introduces a structure where a few fluid antennas can freely switch their locations within a given space \cite{zhu2023Mag}. In this way, the spacing of the available locations (referred to as ``ports'') for fluid antennas can be arbitrarily small. This almost continuously movable feature allows \acp{fas} to keep fluid antennas at favorable positions, thus promising to achieve high diversity and multiplexing gains with very few antennas \cite{wong2020fluid,wong2021fluid,Wenyan'23}. 

Despite these encouraging prospects, the expected gains of \acp{fas} are hard to achieve in practice. In specific, the transmission performance of \acp{fas} heavily relies on the positions of fluid antennas \cite{chai2022port,wong2022fast,Waqar'23}. To ensure favorable antenna placements, the \ac{csi} knowledge of available locations is essential \cite{Skouroumounis'23,ma2023compressed,Rujian'23}. However, the channel estimation for \acp{fas} is challenging. The reason is that, the allowed locations (i.e., the ports) of fluid antennas are densely deployed, leading to very high-dimensional port channels~\cite{zhu2023Mag}. Thereby, it requires an unacceptable number of pilots to acquire the channels. Besides, limited by the hardware structure of \acp{fas}, only a few ports can be connected to \ac{rf} chains for channel measurements within the coherence time, which exacerbates the difficulty of channel estimation. To address the high-dimensional \acp{fas} channels, pilot-reduced channel estimators have been investigated in \cite{Skouroumounis'23,ma2023compressed,Rujian'23}. However, most existing channel estimators rely on some channel assumptions, such as the slow variation \cite{Skouroumounis'23}, angle-domain sparsity \cite{ma2023compressed}, and known angles-of-arrival (AoAs)~\cite{Rujian'23}. When these assumptions are not reasonable, the model mismatch will lead to an unpredictable performance loss.

In this paper, we propose the \ac{sbr} as a general solution to estimate \ac{fas} channels. Different from the existing model-based estimators relying on channel assumptions, the proposed \ac{sbr} builds the experiential kernel of \acp{fas} channels for \ac{csi} acquisition. Specifically, inspired by the Bayesian regression~\cite{williams1995gaussian}, the key idea of \ac{sbr} is to model the \ac{fas} channels as a stochastic process with an experiential kernel, which characterizes the inherent correlation of \ac{fas} channels. Then, the uncertainty of the stochastic process can be successively eliminated by kernel-based sampling and regression. Particularly, the proposed \ac{sbr} is a two-stage scheme. In the first stage, the measured channels are determined by following the principle of maximum posterior variance. In the second stage, the channel measurements are combined with the experiential kernel for process regression.
Then, the mean of the regressed stochastic process is exactly the maximum a posterior (MAP) estimator of \ac{fas} channels. Simulation results reveal that, in both model-mismatched and model-matched cases, the proposed \ac{sbr} can achieve higher estimation accuracy than the existing schemes based on channel assumptions.

The rest of this paper is organized as follows. In Section \ref{sec:model}, the system model of an \ac{fas} is introduced, and the problem of channel estimation is formulated. In Section \ref{sec:SBE}, the general \ac{sbr} is proposed for \ac{fas} channel estimation. In Section \ref{sec:sim}, simulation results are presented to evaluate the estimation performance. Finally, conclusions are drawn in Section \ref{sec:con}.

\textit{Notation:} ${[\cdot]^{-1}}$, ${[\cdot]^{*}}$, ${[\cdot]^{\rm T}}$, and ${[\cdot]^{\rm H}}$ denote the inverse, conjugate, transpose, and conjugate-transpose operations, respectively; ${\bf x}(i)$ denotes the $i$-th entry of vector ${\bf x}$; ${\bf X}({i,j})$, ${\bf X}({j,:})$ and ${\bf X}({:,j})$ denote the $(i,j)$-th entry, the $j$-th row, and the $j$-th column of matrix ${\bf X}$, respectively; ${\rm Tr}(\cdot)$ denotes the trace of its argument; ${\mathsf E}\left(\cdot\right)$ is the expectation of its argument; ${\rm dim}(\cdot)$ is the dimensional of its argument; $\mathcal{C} \mathcal{N}\!\left({\bm \mu}, {\bf \Sigma } \right)$ and $\mathcal{G} \mathcal{P}\!\left({\bm \mu}, {\bf \Sigma } \right)$ respectively denote the complex Gaussian distribution and complex Gaussian process, with mean ${\bm \mu}$ and covariance ${\bf \Sigma }$; $\mathbf{0}_{L}$ is an all-zero vector or matrix with dimension $L$.

\section{System Model}\label{sec:model}

In this paper, we consider the narrowband channel estimation in an uplink \ac{fas}, which consists of an $N$-port \ac{bs} equipped with $M$ fluid antennas and a single-antenna user. The $N$ feeding ports are uniformly distributed along a linear dimension at the receiver. The $M$ fluid antennas can be repositioned to the $M$ locations of $N$ available ports ($M\ll N$), and each antenna is connected to an \ac{rf} chain. Let ${\bf h}\in{\mathbb C}^N$ denote the channels of $N$ ports, and let $P$ denote the number of transmit pilots within a coherence time frame. To characterize the locations of $M$ fluid antennas in timeslot $p$, we introduce the definition of switch matrix as follows:
\begin{definition}[Switch Matrix]
	Binary indicator ${\bf S}_p\in\{0,1\}^{M\times N}$ is defined as the switch matrix of multiple fluid antennas in timeslot $p$. The $(m,n)$-th entry being 1 (or 0) means that the $m$-th antenna is (or not) located at the $n$-th port. Since $M$ of $N$ ports are selected in each timeslot, each row of ${\bf S}_p$ has one entry of 1, and all entries of 1 in ${\bf S}_p$ are not in the same column, i.e., $\|{\bf S}_p({m,:})\|=1$ for all $m\in\{1,\cdots,M\}$,  $\|{\bf S}_p({:,n})\|\in\{0,1\}$ for all $n\in\{1,\cdots,N\}$, and ${\bf S}_p{\bf S}_p^{\rm H} = {\bf I}_M$.
\end{definition}

Utilizing {\bf Definition 1}, the signal vector ${\bf y}_p\in{\mathbb C}^{M}$ received at the \ac{bs} in timeslot $p$ can be modeled as 
\begin{equation}\label{eqn:y_p}
	{\bf y}_p = {\bf S}_p{\bf h}s_p + {\bf z}_p,
\end{equation}
where $s_p$ is the pilot transmitted by the user and ${\bf z}_p \sim \mathcal{C} \mathcal{N}\!\left({\bf 0}_{M},  \sigma^2{\bf I}_{M} \right)$ is the \ac{awgn} at $M$ selected ports. Without loss of generality, we assume that $s_p = 1$ for all $p\in\{1,\cdots,P\}$. Considering the total $P$ timeslots for pilot transmission, we arrive at 
\begin{equation}\label{eqn:y}
	{\bf y} = {\bf S}{\bf h} + {\bf z},
\end{equation}
where ${\bf y} := \left[{\bf y}_1^{\rm T},\cdots,{\bf y}_P^{\rm T}\right]^{\rm T}$, ${\bf S} := \left[{\bf S}_1^{\rm T},\cdots,{\bf S}_P^{\rm T}\right]^{\rm T}$, and ${\bf z} := \left[{\bf z}_1^{\rm T},\cdots,{\bf z}_P^{\rm T}\right]^{\rm T}$. Our goal is to reconstruct the $N$-dimensional channel $\bf h$ according to the $PM$-dimensional noisy pilot ${\bf y}$. Since fluid antennas move almost continuously, $N$ is much larger than $PM$ ($N \gg PM$). Besides, due to the zero-one distribution of ${\bf S}$, most elements of $\bf h$ cannot be observed directly or indirectly. As a result, the channel estimation of \acp{fas} is usually challenging.

\section{Proposed Successive Bayesian Reconstructor}\label{sec:SBE}
In this section, based on the Bayesian regression, we propose the \ac{sbr} as a general solution to realize \ac{fas} channel estimation.  Specifically, in Subsection \ref{subsec:BLR}, the classical Bayesian regression is introduced. Then, in Subsection \ref{subsec:S-BAR}, the proposed \ac{sbr} scheme is illustrated. Finally, in Subsection \ref{subsec:Kernel}, the kernel selection of \ac{sbr} is discussed.

\subsection{Bayesian Regression}\label{subsec:BLR}
Without making any prior assumptions, the attempt to recover the function $f({\bf x})$ from a few samples appears to be a challenging endeavor. Fortunately, by building the experiential kernel of $f({\bf x})$, Bayesian regression can determine the sampling strategy and reconstruct $f({\bf x})$ with a few samples in a non-parametric way. Under this framework, Gaussian process regression (GPR) has become a popular solution \cite{williams1995gaussian}. Specifically, function $f({\bf x})$ can be modeled as a sample of Gaussian process ${\mathcal{GP}}\left(\mu\left({\bf x}\right), k\left({\bf x},{\bf x}'\right)\right)$. It is completely specified by its mean $\mu\left({\bf x}\right)$ and its kernel $k\left({\bf x},{\bf x}'\right)$, which encodes the smoothness of regressed $f({\bf x})$. In timeslot $t$, consider a prior ${\mathcal{GP}}\left(\mu\left({\bf x}\right), k\left({\bf x},{\bf x}'\right)\right)$ over $f({\bf x})$. Let ${\bm \gamma}^t := [\gamma^1,\cdots,\gamma^t]^{\rm T}$ denote $t$ noisy measurements for points in ${\cal A}^t := \{{\bf x}^1,\cdots,{\bf x}^t\}$, where $\gamma^i = f({\bf x}^i) + n_i$ with $n_i\sim{\cal{CN}}\left(0, \delta^2\right)$. It is easy to prove that, given ${\bm \gamma}^t$, the posterior over $f({\bf x})$ is also a Gaussian process whose mean and covariance are
\begin{align}
	\label{eqn:mu_x}
	{\mu^t\left({\bf x}\right)} &= \mu\left({\bf x}\right) + \left({\bf k}^t({\bf x})\right)^{\rm H}\! \left( {\bf K}^t + \delta^2{\bf I}_t \right)^{-1}\!\left({\bm \gamma}-{\bm \mu}^t\right), \\ 
	\label{eqn:k_x_x'}
	{k^t\left({\bf x},{\bf x}'\right)} &= k\left({\bf x},{\bf x}'\right) - \left({\bf k}^t({\bf x})\right)^{\rm H} \! \left( {\bf K}^t + \delta^2{\bf I}_t \right)^{-1}\!{\bf k}^t({\bf x}'),
\end{align}
where ${\bf k}^t({\bf x}): = \left[k\left({\bf x}^1,{\bf x}\right),\cdots,k\left({\bf x}^t,{\bf x}\right)\right]^{\rm T}$; ${\bm \mu}^t: = \left[\mu\left({\bf x}^1\right),\cdots,\mu\left({\bf x}^t\right)\right]^{\rm T}$
; and the $(i,j)$-th entry of ${\bf K}^t\in{\mathbb C}^{t\times t}$ is $k\left({\bf x}^i,{\bf x}^j\right)$, for all $i,j\in\{1,\cdots,t\}$.

Then, the next candidate point to be sampled, i.e., ${\bf x}^{t+1}$, can be determined based on the updated posterior. For successive sampling, sampling the point with the maximum posterior variance can obtain the most information. By assuming that ${\bf x}\in {\cal S}$, ${\bf x}^{t+1}$ can be chosen according to
\begin{align}
	\label{eqn:x_t+1}
	{\bf x}^{t+1} = \mathop {\arg \max }\limits_{{\bf x}\in {\cal S}/{\cal A}^{t}} ~ {k^{t}\left({\bf x},{\bf x}\right)},
\end{align}
where $/$ is the set difference. By letting $t \to \infty$, the value of variance $k^t\left({\bf x},{\bf x}\right)$ decreases asymptotically, which means that the uncertainty of $f({\bf x})$ is reduced. After reaching the tolerance threshold, the posterior mean ${\mu^t\left({\bf x}\right)}$ can be viewed as a MAP estimator of $f({\bf x})$ \cite{williams1995gaussian}.

\subsection{Proposed \ac{sbr} Scheme}\label{subsec:S-BAR}
In each pilot timeslot, $M$ fluid antennas move positions and measure channels, thus the channel estimation of \acp{fas} is similar to a successive sampling process. Since the port spacing is short, the \ac{fas} channels are highly correlated. These features inspire us to recover $\bf h$ through Bayesian regression. To reconstruct \ac{fas} channels based on experiential kernel, we model $\bf h$ as a sample of Gaussian process ${\mathcal{GP}}\left({\bf 0}_N, {\bm \Sigma}\right)$. Semidefinite Hermitian matrix ${\bm \Sigma}\in{\mathbb C}^{N\times N}$ is called the kernel or prior covariance, of which the selection will be introduced in Subsection \ref{subsec:Kernel}. Then, the proposed \ac{sbr} scheme is summarized in {\bf Algorithm 1}. For clarity, the basic principle of \ac{sbr} is firstly introduced as follows. 

\begin{algorithm}[!t] 
	\caption{Proposed Successive Bayesian Reconstructor} 
	\begin{algorithmic}[1]
		\REQUIRE  
		Number of pilots $P$, kernel ${\bm \Sigma}$.
		\ENSURE 
		Reconstructed \ac{fas} channel $\hat{\bf h}$. 
		\STATE {\it \# Stage 1 (Offline Design):} 
		\STATE Initialization: $\Omega = \varnothing$, ${\bf S}_p = {\bf 0}_{M\times N}$ for all $p\in\{1,\cdots,P\}$ 
		\FOR{$p\in\{1,\cdots,P\}$}
		\FOR{$m\in\{1,\cdots,M\}$}
		\STATE Posterior covariance update: Calculate ${\bm \Sigma }_{\Omega}$ by (\ref{eqn:Sigma_O})
		\STATE Candidate selection: $n^\star = \mathop {\arg \max }\limits_{n\in \{1,\cdots,N\}/\Omega} ~ {{\bm \Sigma }_{\Omega}(n,n)}$
		\STATE Switch matrix update: ${\bf S}_p(m,n^\star) = 1$
		\STATE Sequence update: $\Omega = \Omega \cup \{n^\star\}$
		\ENDFOR
		\ENDFOR
		\STATE Merge switch matrices: ${\bf S} := \left[{\bf S}_1^{\rm T},\cdots,{\bf S}_P^{\rm T}\right]^{\rm T}$
		\STATE Weight calculation: ${\bf w} = ( {\bm \Sigma}(\Omega,\Omega) + \sigma^2{\bf I}_{PM} )^{-1} {\bm \Sigma}(\Omega,:)$
		\STATE {\it \# Stage 2 (Online Regression):}
		\STATE Employ the designed switch matrix ${\bf S}$ at the \ac{bs}, and then obtain the received pilot: ${\bf y} = {\bf S}{\bf h} + {\bf z}$
		\STATE Channel reconstruction: $\hat{\bf h} = {\bf w}^{\rm H}{\bf y}$
		\RETURN Reconstructed \ac{fas} channel $\hat{\bf h}$ 
	\end{algorithmic}
\end{algorithm}
\subsubsection{Algorithmic Principle}

At some moment, let $\Omega$ denote the index sequence of the measured channels and let ${\bf y}_\Omega \in {\mathbb C}^{{\rm dim}(\Omega)}$ denote the corresponding received pilots, which is from ${{\bf{y}}_\Omega } = {\bf{h}}\left( \Omega  \right) + {{\bf{z}}_\Omega }$ with ${\bf{z}}_\Omega\sim{\cal CN}({\bf 0}_{{\rm dim}(\Omega)},\sigma^2{\bf I}_{{\rm dim}(\Omega) })$ being the \ac{awgn}. For given ${{\bf{y}}_\Omega }$, the posterior mean ${\bm \mu }_{\Omega}$ and posterior covariance ${\bm \Sigma }_{\Omega}$ of $\bf h$ can be calculated by:
\begin{align}
	\label{eqn:mu_O}
	{\bm \mu }_{\Omega} &= {\bm \Sigma}(:,\Omega) \left( {\bm \Sigma}(\Omega,\Omega) + \sigma^2{\bf I}_{{\rm dim}(\Omega)} \right)^{-1}{\bf y}_\Omega, \\
	\label{eqn:Sigma_O}
	{\bm \Sigma }_{\Omega} &= {\bm \Sigma } -  \left( {\bm \Sigma}(\Omega,:) \right)^{\rm H}\left( {\bm \Sigma}(\Omega,\Omega) + \sigma^2{\bf I}_{{\rm dim}(\Omega)} \right)^{-1}{\bm \Sigma}(\Omega,:).
\end{align}
For given $\Omega$, the next candidate channel to be measured can be determined by finding the index associated with the largest posterior variance, i.e.,
\begin{align}
	\label{eqn:n_O}
	n^\star = \mathop {\arg \max }\limits_{n\in \{1,\cdots,N\}/\Omega} ~ {{\bm \Sigma }_{\Omega}(n,n)}.
\end{align}
Subsequently, we can update $\Omega$ by $\Omega \cup \{n^\star\}$ and repeat the above process until the posterior mean ${\bm \mu }_{\Omega}$ can well approximate $\bf h$.

\subsubsection{Observations}
From the above equations, we obtain the following three observations.
\begin{itemize}
	\item (\ref{eqn:mu_O}) indicates that, the posterior mean ${\bm \mu }_{\Omega}$ is the linear weighted sum of pilot ${\bf y}_\Omega$, i.e., ${\bm \mu }_{\Omega}={\bf w}^{\rm H}{\bf y}_\Omega$, wherein the weight ${\bf w}:=\left( {\bm \Sigma}(\Omega,\Omega) + \sigma^2{\bf I}_{{\rm dim}(\Omega)} \right)^{-1} {\bm \Sigma}(\Omega,:)$ only relies on the kernel ${\bm \Sigma}$.
	\item (\ref{eqn:Sigma_O}) shows that posterior covariance ${\bm \Sigma }_{\Omega}$ only relies on kernel ${\bm \Sigma}$ and is unrelated to the received pilot ${\bf y}_\Omega$.
	\item (\ref{eqn:n_O}) suggests that the next channel to be measured only relies on the posterior covariance ${\bm \Sigma }_{\Omega}$. 
\end{itemize}
These observations reveal that, the switch matrix $\bf S$ and the weight $\bf w$ are unrelated to the received pilot $\bf y$, thus they can be designed offline and then deployed online to reduce the complexity. Thereby, the proposed \ac{sbr} can be realized in the following two stages.

\subsubsection{Stage 1 (Offline Design)}
Since index sequence $\Omega$ is determined by the posterior covariance ${\bm \Sigma }_{\Omega}$, and ${\bm \Sigma }_{\Omega}$ only relies on the kernel ${\bm \Sigma}$. The switch matrix ${\bf S}\in{\{0,1\}}^{PM\times N}$ and the weight ${\bf w}\in{\mathbb C}^{PM}$ for recovering ${\bf h}\in{\mathbb C}^N$ can be designed offline at the first stage. By updating ${\bm \Sigma }_{\Omega}$ in (\ref{eqn:Sigma_O}) and $n^\star$ in (\ref{eqn:n_O}) alternatingly until ${\rm dim}(\Omega)=PM$, sequence $\Omega$ can collect all required indexes of the $M$ selected ports in $P$ pilot timeslots. 

Then, recall that we have ${\bf{h}}\left( \Omega  \right) = {\bf S}{\bf{h}}$. To achieve the conversion from $\Omega$ to ${\bf S}$, we can initialize $\bf S$ as an all-zero matrix and then fill in an one at the position associated with the selected index in each of its row. Note that, this operation naturally satisfies $\|{\bf S}({m,:})\|=1$ for all $m\in\{1,\cdots,M\}$,  $\|{\bf S}({:,n})\|\in\{0,1\}$ for all $n\in\{1,\cdots,N\}$, and ${\bf S}{\bf S}^{\rm H} = {\bf I}_{PM}$. These properties ensure that the designed $\bf S$ is practically implementable in \acp{fas}. After obtaining $\Omega$, the weight for reconstructing $\bf h$ can be obtained by 
\begin{equation}
	{\bf w}=\left( {\bm \Sigma}(\Omega,\Omega) + \sigma^2{\bf I}_{PM} \right)^{-1} {\bm \Sigma}(\Omega,:).
\end{equation}
\subsubsection{Stage 2 (Online Regression)}
Since {\it Stage 1} is realized offline, the switch matrix ${\bf S}$ and weight ${\bf w}$ can be designed and saved at the \ac{bs} in advance. In {\it Stage 2}, the scheme is then employed online for channel measurements. The $M$ fluid antennas of the \ac{bs} will move and receive pilots according to the designed ${\bf S}$, arriving at the noisy pilot $\bf y$. Finally, according to the MAP estimator in (\ref{eqn:mu_O}), channel $\bf h$ can be reconstructed by $\hat{\bf h} = {\bf w}^{\rm H}{\bf y}$, which completes the proposed \ac{sbr}.

\subsubsection{Computational Complexity}
The proposed \ac{sbr} incorporates a hybrid offline and online implementation process, thereby substantially reducing its computational complexity in practical applications. Specifically, the signal processing of \ac{sbr} is composed of two stages. In {\it Stage 1}, the computational complexity is dominated by the calculation of posterior covariance ${\bm \Sigma }_{\Omega}$, which is updated $PM$ times. According to (\ref{eqn:Sigma_O}), the complexity of {\it Stage 1} is ${\cal O}\left({P^2}{M^2}\left( {{P^2}{M^2} + NPM + {N^2}} \right)\right)$.  In {\it Stage 2}, the computational complexity is from the weighted sum of received pilot $\bf y$, i.e., $\hat{\bf h} = {\bf w}^{\rm H}{\bf y}$, thus the computational complexity is ${\cal O}\left(N\right)$. Note that, although the complexity of {\it Stage 1} is high, {\it Stage 1} can be implemented offline in advance. From the perspective of practical employment, the effective complexity of \ac{sbr} scheme is only linear to the number of ports $N$.

\subsection{Kernel Selection for \ac{sbr} Scheme}\label{subsec:Kernel}
The selection of kernel $\bm \Sigma$ determines the shape and flexibility of the proposed \ac{sbr}, which in turn affects its ability to capture patterns and make accurate reconstruction. Considering the localized correlation property of \ac{fas} channels, an appropriate kernel should assign higher similarity to nearby ports and decrease influence rapidly with distance. Let ${\bf x}_n$ denote the position of the $n$-th port. Three kernel selections are recommended as follows.

\subsubsection{Exponential Kernel}
The exponential kernel ${\bm \Sigma}_{\rm exp} $ is a popular choice in regression, given by
\begin{equation}
	{\bm \Sigma}_{\rm exp}(n,n') = \alpha^2  e^{-\frac{\|{\bf x}_n-{\bf x}_{n'}\|^2}{\eta^2}}
\end{equation}
for all $n,n'\in\{1,\cdots,N\}$, where $\alpha$ and $\eta$ are adjustable hyperparameters. Compared with the other kernels, the exponential kernel is less sensitive to outliers, which makes it suitable to recover channels without obvious regularity.

\subsubsection{Bessel Kernel}
The Bessel kernel ${\bm \Sigma}_{\rm bes}$ is well-suited for capturing and modeling complex-valued data with oscillatory or periodic patterns, given by
\begin{equation}
	{\bm \Sigma}_{\rm bes}(n,n') = \alpha^2 J_v\left(\frac{\|{\bf x}_n-{\bf x}_{n'}\|}{\eta}\right)
\end{equation}
for all $n,n'\in\{1,\cdots,N\}$, wherein $J_v$ is the $v$-order Bessel function of the first kind. It has the flexibility to adapt to data that exhibits regular and repeating fluctuations, thus ${\bm \Sigma}_{\rm bes}$ is suitable to reconstruct the channels with periodic patterns.
\subsubsection{Covariance Kernel}
An ideal approach is to use the real covariance of $\bf h$ as the kernel for reconstruction, i.e., ${\bm \Sigma}_{\rm cov}={\mathsf E}\left({\bf h}{\bf h}^{\rm H}\right)$. Since ${\bm \Sigma}_{\rm cov}$ is unknown in practice, we can train an approximated ${\bm \Sigma}_{\rm cov}$ before employing \ac{sbr}, given by
\begin{equation}\label{eqn:cov_ker}
	{\bm \Sigma}_{\rm cov} \approx \frac{1}{T}\sum\limits_{t = 1}^T {{{\bf{h}}_t}{\bf{h}}_t^{\rm H}},
\end{equation}
where ${\bf{h}}_t$ is the channel at the $t$-th training timeslot and $T$ is the number of training timeslots. Since the channel covariance ${\mathsf E}\left({\bf h}{\bf h}^{\rm H}\right)$ does not change so frequently as channels, ${\bm \Sigma}_{\rm cov}$ is only updated in a large timescale. 

\section{Simulation Results}\label{sec:sim}
In this section, simulation results are provided to verify the effectiveness of the proposed \ac{sbr} scheme. Since we have assumed the normalized transmit power, the receiver \ac{snr} is defined as ${\rm SNR}=\frac{{\mathsf E}\left(\|{\bf h}\|^2\right)}{\sigma^2}$, of which the default value is set to $20$ dB. Let $\hat{\bf h}$ denote the estimated value of channel $\bf h$. The performance is evaluated by the \ac{nmse}, i.e., ${\rm NMSE}={\mathsf E}\left(\frac{\|{\bf h}-\hat{\bf h}\|^2}{\|{\bf h}\|^2}\right)$.

\begin{figure*}[!t]
	\centering
	\hspace{-5mm}
	\subfigure{\includegraphics[width=0.95\textwidth]{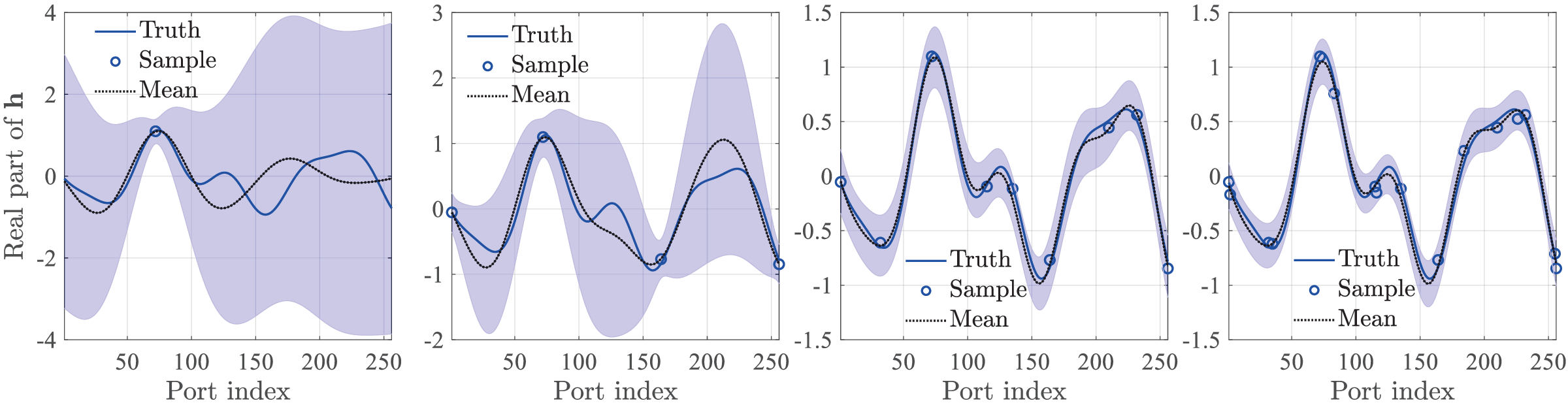}}
	\\
	\vspace{-3mm}
\begin{flushleft}
	{\footnotesize \quad\quad\quad\quad\quad\quad(a) $P=1$, $M=1$.\quad\quad\quad\quad\quad\quad\quad(b) $P=2$, $M=2$.\quad\quad\quad\quad\quad\quad\quad(c) $P=3$, $M=3$.\quad\quad\quad\quad\quad\quad\quad(d) $P=4$, $M=4$.}
\end{flushleft}
	\vspace{-1mm}
	\hspace{-5mm}	
	\subfigure{\includegraphics[width=0.95\textwidth]{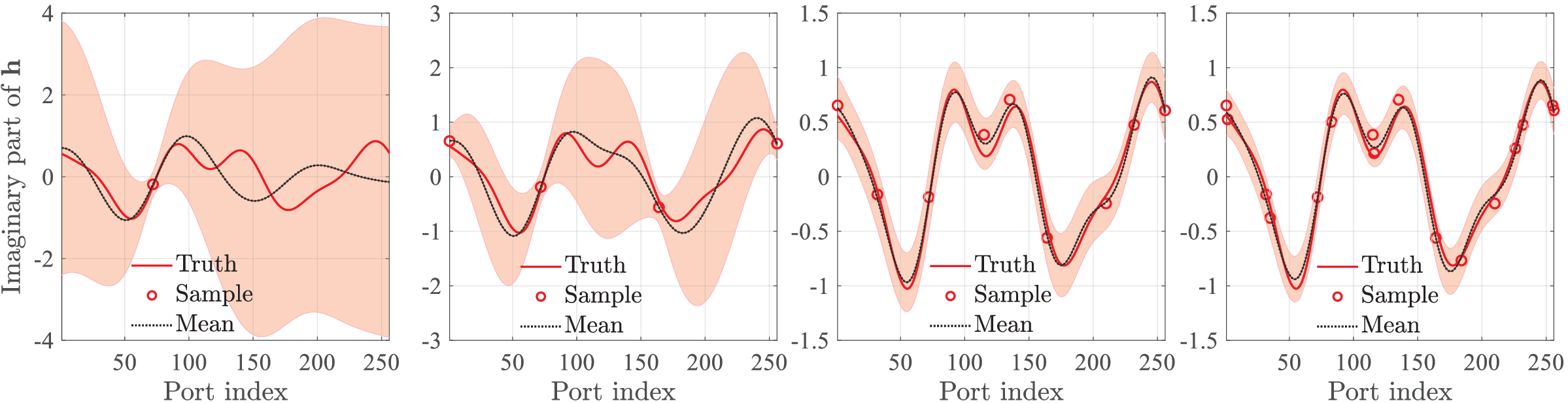}}
	\vspace{-3mm}
	\begin{flushleft}
		{\footnotesize \quad\quad\quad\quad\quad\quad(e) $P=1$, $M=1$.\quad\quad\quad\quad\quad\quad\quad(f) $P=2$, $M=2$.\quad\quad\quad\quad\quad\quad\quad(g) $P=3$, $M=3$.\quad\quad\quad\quad\quad\quad\quad(h) $P=4$, $M=4$.}
	\end{flushleft}
	\vspace*{-1em}
	\caption{ An illustration of employing \ac{sbr} scheme to estimate \ac{fas} channel $\bf h$. (a)-(d) provide the real part of $\bf h$ versus the index of ports. (e)-(h) provide the imaginary part of $\bf h$ versus the index of ports. Particularly, the curve ``Truth'' denotes the real channel $\bf h$, and the circle marks denote the sampled (measured) channels. The dotted line ``Mean'' denotes the posterior mean of Bayesian regression ${\bm \mu }_{\Omega}$, i.e., the estimated channel $\hat{\bf h}$.  The highlighted shadows in the figures represent the confidence intervals of $\bf h$, defined as $[{\bm \mu }_{\Omega}(n)-3{{\bm \Sigma }_{\Omega}(n,n)},{\bm \mu }_{\Omega}(n)+3{{\bm \Sigma }_{\Omega}(n,n)}]$ for the $n$-th port.
	}
	\label{img:GPR}
	\vspace{-1em}
\end{figure*}

\subsubsection{Simulation Setup}
The simulations are provided based on both the QuaDRiGa channel model and the spatially-sparse clustered (SSC) channel model. For existing model-based estimators, these two models can be viewed as the matched case and mismatched case, respectively. Otherwise particularly specified, the system parameters are set as: $N=256$, $M=4$, $P=10$. The carrier frequency is set to $f_c = 3.5$ GHz, and the length of the fluid antenna array is set to $W=10\lambda$. For the QuaDRiGa channel model, all parameters are generated according to Table 7.7.1-2 in 3GPP TR 38.901. For the SSC channel model, the number of clusters is set to $C=9$ and that of rays is set to $R=100$. Both models have assumed the maximum angle spread to be 5$^\circ$. For kernel settings, the hyperparameters are set as $\alpha=1$ and $\eta=\sqrt{\frac{\lambda}{2\pi}}$ to generate the exponential kernel ${\bm \Sigma}_{\rm exp}$ and Bessel kernel ${\bm \Sigma}_{\rm bes}$ \cite{williams1995gaussian}. Inspired by the covariance model in \cite{wong2021fluid,wong2020fluid}, the order of Bessel function in ${\bm \Sigma}_{\rm bes}$ is set to $\nu = 0$. To account for an ideal baseline, the number of training timeslots is set to $T=100$ to train the covariance kernel ${\bm \Sigma}_{\rm cov}$.

\subsubsection{Simulation Schemes}
We consider the following three schemes for simulations. {\bf 1) FAS-OMP:} Assuming that the \ac{fas} channels are spatially sparse, the scheme in \cite{ma2023compressed} is modified and employed at the \ac{bs} to explicitly estimate $\bf h$. {\bf 2) SeLMMSE: } The SeLMMSE proposed in \cite{Skouroumounis'23} is adopted to estimate channel $\bf h$, which can be achieved by sequentially measuring channels of $PM$ equally-spaced ports and then using zero-order interpolation to reconstruct $\bf h$. {\bf 3) Proposed \ac{sbr}:} Given a kernel $\bm \Sigma$, the proposed \ac{sbr} scheme, i.e., {\bf Algorithm 1}, is employed to estimate $\bf h$. Particularly, due to the lack of obvious regularity, the exponential kernel ${\bm \Sigma}_{\rm exp}$ is selected as the input of \ac{sbr} for QuaDRiGa channels. Due to their periodic patterns in the spatial domain, the Bessel kernel ${\bm \Sigma}_{\rm bes}$ is selected as the input of \ac{sbr} for SSC channels. To provide an ideal baseline, the pre-trained covariance kernel ${\bm \Sigma}_{\rm cov}$ is considered for both channel models.

\subsubsection{Simulation Results}

\begin{figure}[!t]
	\centering
	\includegraphics[width=0.43\textwidth]{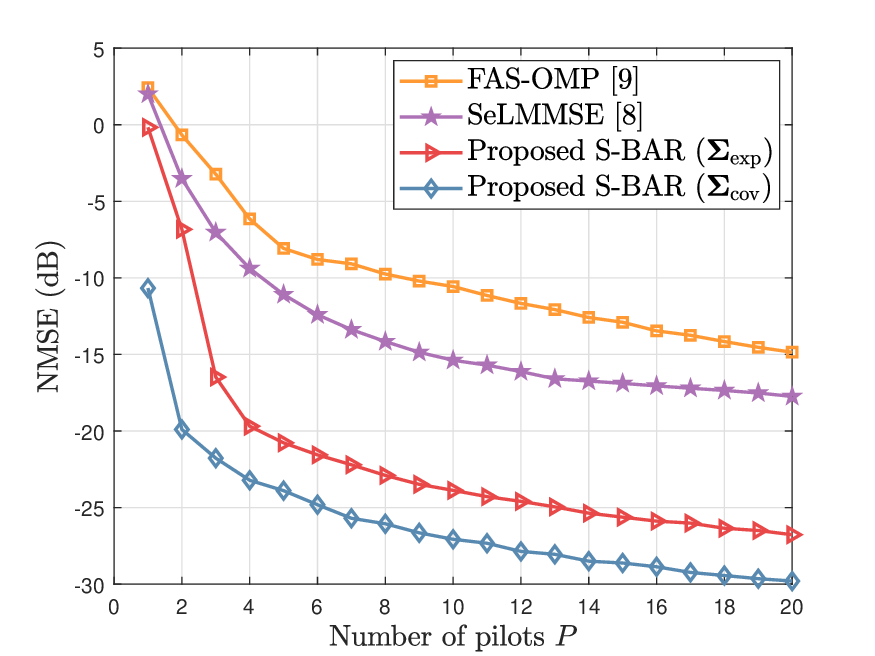}
	\vspace*{-1em}
	\caption{Model-mismatched case: The \ac{nmse} as a function of the number of pilots $P$ under the assumption of QuaDRiGa channel model.}
	\vspace*{-1em}
	\label{img:NMSE_vs_P}
\end{figure}

To better understand the working principle of the proposed \ac{sbr}, we plot Fig. \ref{img:GPR} to intuitively show its behavior, where the QuaDRiGa channel model is considered and the covariance kernel ${\bm \Sigma}_{\rm cov}$ is used to enable \ac{sbr}. From this figure, we have two observations. Firstly, as the number of samples increases, the confidence interval is gradually reduced. It indicates that more pilots or antennas can better eliminate the uncertainty of \ac{fas} channels. Secondly, one can note that the sample spacing is usually large. The reason is that, for each sampling, the proposed \ac{sbr} samples the channel with the largest posterior variance. When a port is selected and measured, the channel uncertainty of its nearby ports will decrease, which reduces the trend of selecting them as samples. 

\begin{figure}[!t]
	\centering
	\includegraphics[width=0.43\textwidth]{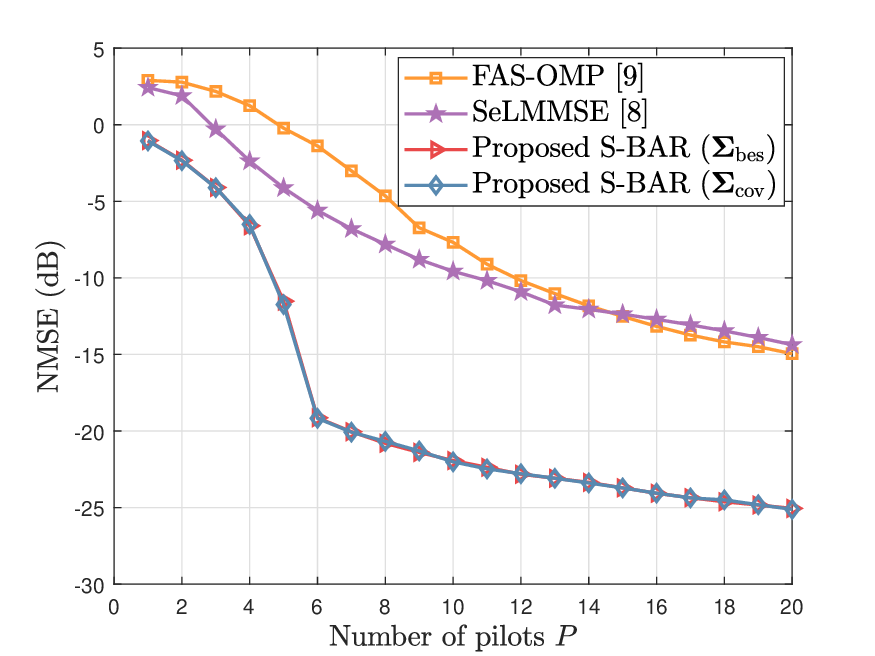}
	\vspace*{-1em}
	\caption{Model-matched case: The \ac{nmse} as a function of the number of pilots $P$ under the assumption of SSC channel model.}
	\vspace*{-1em}
	\label{img:NMSE_vs_P_SV}
\end{figure}

Then, we plot the \ac{nmse} as a function of the number of pilots $P$ in Fig. \ref{img:NMSE_vs_P} for QuaDRiGa model and Fig. \ref{img:NMSE_vs_P_SV} for SSC model, respectively. From these two figures, we have the following observations. 
Firstly, the proposed \ac{sbr} achieves the highest estimation accuracy in both cases. The reason is that, the existing methods do not fully utilize the channel prior for estimation. For FAS-OMP, due to the non-ideal port selection, the information provided by the randomly measured channels may not be sufficient to capture all channel patterns. For SeLMMSE, the unmeasured channels are directly obtained by zero-order interpolation, while their potential estimation errors are not considered. In contrast, the proposed \ac{sbr} incorporates the effect of prior correlation into its estimator, which naturally considers the potential estimation errors of all channels. Through kernel-based sampling and regression, \ac{sbr} can eliminate the uncertainty of many channels with a few pilots. Secondly, the \ac{sbr} enabled by the experiential kernels ${\bm \Sigma}_{\rm exp}$ and ${\bm \Sigma}_{\rm bes}$ can achieve similar performance as that enabled by covariance kernel ${\bm \Sigma}_{\rm cov}$. Recall that ${\bm \Sigma}_{\rm exp}$ and ${\bm \Sigma}_{\rm bes}$ are generated by experiential parameters, while ${\bm \Sigma}_{\rm cov}$ is trained from real channel data. This observation indicates that, even if the real channel covariance ${\mathsf E}\left({\bf h}{\bf h}^{\rm H}\right)$ is unknown, experiential parameters still allow \ac{sbr} to achieve considerable performance. 

\section{Conclusions}\label{sec:con}
In this paper, we have proposed \ac{sbr} as a general solution to estimate channels in \acp{fas}. Different from the existing channel estimators relying on channel assumptions, the general \ac{sbr} utilizes the experiential kernel to acquire \ac{csi} in a non-parametric way. Inspired by the Bayesian regression, the proposed \ac{sbr} can select a few informative channels for measurement and combine them with experiential kernel to reconstruct high-dimensional \ac{fas} channels. Simulation results reveal that, in both model-mismatched and model-matched cases, the proposed \ac{sbr} can achieve much higher estimation accuracy than the existing schemes. 

\section*{Acknowledgment}
This work was supported in part by the National Natural Science
Foundation of China (Grant No. 62031019).

\footnotesize
\bibliographystyle{IEEEtran}
	
\bibliography{IEEEabrv,reference}
	
\end{document}